\documentclass{elsart3}

\usepackage{epsfig}

\usepackage{amssymb}

\begin{document}

\begin{frontmatter}



\title{Competition of Orbital Antiferromagnetism With Q-Triplet-Pairing in the
Ferromagnetic Superconductor UGe$_2$}


\author[label1]{G. Varelogiannis} \author[label2]{P. Thalmeier} 
\author[label1]{S. Tsonis}

\address[label1]{Department. of Physics, National Technical University of
Athens, GR-15780 Athens, Greece}

\address[label2]{Max-Planck-Institute for Chemical Physics of Solids,
D-01187 Dresden, Germany}

\begin{abstract}
Within a multicomponent BCS-like framework we study the coexistence and
competition of various CDW and SC order parameters in the presence of 
a strong ferromagnetic background. We find that the competition
of unconventional CDW, also called orbital antiferromagnet, with SC
at finite momentum (${\bf Q}$-triplet pairing) shows unique characteristics
like an extreme sensitivity on the deviations from nesting. 
We argue that pressure in UGe$_2$ creates deviations from the nesting
and report a phase diagram in qualitative agreement with the observed behavior
of UGe$_2$.

\end{abstract}

\end{frontmatter}


The discovery of superconductivity (SC) well inside the 
itinerant ferromagnetic (FM) state of UGe$_2$ \cite{Saxena} is a true challenge
in the theory of superconductivity in strongly correlated systems. It
appears now quite clear that
SC develops only inside the FM state quite suddenly when
the critical temperature $T^*$ of a hidden order state
inside the FM state is reduced towards zero by applying pressure. 
The existence of a true thermodynamic phase with hidden order below
T$^*$ has been suggested by specific heat experiments
\cite{Tateiwa}. The FM transition is apparently of first order
\cite{Pfleiderer}  which is very hard to reconcile with any approach
that would associate superconductivity with the fluctuations of
the FM or the $T^*$ hidden order as in the quantum critical point scenario.

Within our model we obtain a pressure- temperature phase diagram that
displays many of the observed characteristics of UGe$_2$. We propose
that $T^*$ corresponds to a CDW formation and we have studied the
competition of various CDW and SC order parameter symmetries
within a BCS like mean-field framework in the
presence of a fully polarized FM background, i.e. frozen spins.  
We use a Nambu type formalism with spinors defined by
$$
\Psi^{\dagger}_{\bf{k}}=\bigl(c^{\dagger}_{\bf{k}},
c_{-\bf{k}},c^{\dagger}_{\bf{k}+\bf{Q}},c_{-\bf{k}-\bf{Q}}\bigr)
\eqno(1)
$$
Nesting in the fully polarized band is responsible for the
CDW transition that we associate with the $T^*$ line in UGe$_2$.
Pressure reduces $T^*$ because it removes the nesting features.
To discuss this phenomenon we write the electron
dispersion $\xi_{\bf k}$ as a sum of
particle-hole symmetric terms reponsible for nesting
and particle-hole asymmetric terms that represent the deviations
from nesting:
$\xi_{\bf k}=\gamma_{\bf k}+\delta_{\bf k}$ where
$2\gamma_{\bf k}=\xi_{\bf k}-\xi_{\bf{k}+\bf{Q}}$ and
$2\delta_{\bf k}=\xi_{\bf k}+\xi_{\bf{k}+\bf{Q}}$.
When $\delta_{\bf k}=0$ there is particle-hole symmetry or perfect
nesting at the wavevector $\bf{Q}$.

Our approach allows to study
SC and CDW states which transform either odd or even
under ${\bf k}\rightarrow -{\bf k}$ (inversion) or
${\bf k}\rightarrow {\bf k}+{\bf Q}$.
The possible superconducting states must be odd under inversion 
because the spins are frozen. We therefore examine a special type
of unitary triplet SC state. Our analysis shows that
with these symmetry constraints we must necessarily have
$\Delta^{\bf Q}_{\bf{k}+\bf{Q}}=-\Delta^{\bf Q}_{\bf k}$
where $\Delta^{\bf Q}_{\bf k}$ is the SC gap for pairing at finite
momentum ${\bf Q}$
also called ${\bf Q}$-triplet pairing or $\pi$-triplet pairing.
For the zero momentum pairing state the order parameter
$\Delta_{\bf k}$ must be purely imaginary.
Concerning the CDW states, for states with both even and odd parity
the order parameter
$W_{\bf k}$ is real if $W_{\bf k}=W_{\bf{k}+\bf{Q}}$
and imaginary if $W_{\bf k}=-W_{\bf{k}+\bf{Q}}$.
The CDW states which obey $W_{\bf k}=-W_{\bf{k}+\bf{Q}}$
are unconventional CDW states (UCDW) also called orbital
antiferromagnetic states due to the staggered current pattern
they induce on the lattice plaquettes.

A study of the coexistence and
competition of all accessible SC and CDW states 
will be presented elsewhere. Here we focus on a single case which
is perhaps
the most exciting one showing unique characteristics
that may plausibly correspond to the physical situation observed
in the FM state of UGe$_{2}$.
It corresponds to the competition of the even parity UCDW state 
with the finite-${\bf Q}$ SC state.
To simplify the notation from now on we will note $\Delta^{\bf Q}_{\bf k}\equiv
\Delta_{\bf k}$.
The coupled gap equations that we obtain in that case have
the following form:
$$
W_{\bf k}=\sum_{\bf k'}
V^{CDW}_{k,k'}W_{\bf k'}{1\over 4
\sqrt{\gamma^2_{\bf k'}+W^2_{\bf k'}}}
\times
$$
$$
\times
\Bigl( \tanh
{
E_{+}({\bf k})\over
2T}+
\tanh
{
E_{-}({\bf k})\over
2T}\Bigr)
\eqno(2)
$$
$$
\Delta_{\bf k}=
\sum_{\bf k'}
V^{SC}_{k,k'} \Delta_{\bf k'}
{1\over 4
\sqrt{\delta^2_{\bf k'}+\Delta^2_{\bf k'}}}
\times
$$
$$
\times
\Bigl( \tanh
{
E_{+}({\bf k})\over
2T}-
\tanh
{
E_{-}({\bf k})\over
2T}\Bigr)
\eqno(3)
$$
where $W_{\bf k}$ is the UCDW gap and
$$
E_{\pm}({\bf k})=
\sqrt{\gamma^2_{\bf k}+W^2_{\bf k}}\pm
\sqrt{\delta^2_{\bf k}+\Delta^2_{\bf k}}
\eqno(4)
$$
The system of gap equations (2) and (3) is very peculiar
showing a unique influence of particle-hole asymmetry on the 
SC versus UCDW competition. Indeed, in the
UCDW gap equation (2), in the zero temperature regime,
only the particle-hole symmetric terms $\gamma_{\bf k}$ of the
dispersion enter. In the SC gap equation (3) instead,
in the same temperature regime, the particle-hole asymmetric terms
$\delta_{\bf k}$ appear. On the other hand at finite temperatures both terms
enter in both gap equations. 

\begin{figure}[h]
\centerline{\psfig{figure=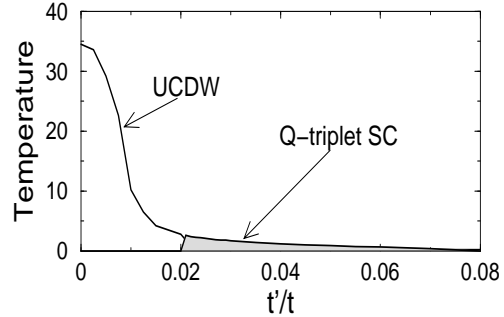,width=6.5cm,height=4.5cm,angle=0}}
\vspace{0.0cm}
\caption{Phase diagram showing the competition of UCDW with Q-triplet SC
as a function of the particle-hole asymmetry. The two states
do not coexist but rather exclude each other.}
\label{FIG1}
\end{figure}

We solve numerically the gap equations (2) and (3)
considering a n.n.n tight binding model for the majority spin band of
UGe$_2$. In which case $\gamma_{\bf k} = t\bigl( \cos k_x+\cos
k_y\bigr)$ and $\delta_{\bf k}=t'\cos k_x\cos k_y$ 

We assume that the effect of pressure is to create deviations
from nesting by enhancing the ratio $t'/t$.
Some of our results are shown in Fig.~\ref{FIG1}.
Bearing in mind that the relationship of pressure with $t'/t$ 
is likely nonlinear, the results shown in  Fig.~\ref{FIG1} are
quite similar to the pressure behaviour of UGe$_2$. The UCDW gap
is very rapidly reduced with particle hole asymmetry and above a
critical $t'/t$ (i.e. a critical pressure) Q-triplet SC suddenly
appears. The fact that SC is
taking its maximal critical temperature at the point where the UCDW disappears 
is not related to any role played by the fluctuations of the incipient
UCDW order parameter since the latter is in rather strong competition with SC.

\end{document}